\documentclass[prb,preprint]{revtex4-1} 

\usepackage{amsmath}  
\usepackage{amsfonts} 
\usepackage{graphicx} 
\usepackage{comment}

\usepackage{color}
\definecolor{ndcol}{rgb}{0.75,0.0,0.0}
\definecolor{myred}{rgb}{0.8,0.0,0.0}
\definecolor{myblue}{rgb}{0.0,0.0,0.8}
\definecolor{mydkblue}{rgb}{0.2,0.2,0.7}
\definecolor{mygreen}{rgb}{0.0,0.4,0.0}
\definecolor{mypurple}{rgb}{0.45,0.09,0.63}
\definecolor{mybrown}{rgb}{0.5,0.1,0.098}
\definecolor{mygray}{rgb}{0.2,0.2,0.2}

\newcommand*{\jared}[1]{{\color{myblue}#1}}                                          

\begin{document}

\title{Problem Roulette: Studying Introductory Physics in the Cloud } 

\author{August E. Evrard}
\email{evrard@umich.edu} 
\author{Michael Mills}
\email{mcmills@umich.edu} 
\author{David Winn}
\author{Kathryn Jones}
\author{Jared Tritz}
\author{Timothy A. McKay}

\affiliation{Department of Physics, University of Michigan, Ann Arbor, MI 48109-1040}

\date{\today}

\begin{abstract} 
Problem Roulette (PR) is a web-based study service at the University of Michigan that 
offers random-within-topic access to a large library of past exam problems in 
introductory physics courses.
Built on public-private cloud computing infrastructure, PR served nearly 1000 students during Fall 2012 term, delivering more than 60,000 problem pages.  The service complements that of 
commercial publishing houses by offering problems authored by local professors.   
We describe the service architecture, including reporting and analytical capabilities, 
and present an initial evaluation of the impact of its use.
Among roughly 500 students studying  electromagnetism, we find that the 229 
students who worked fifty or more problems over the term outperformed their 
complement by 0.40 grade points (on a 4.0 scale).
This improvement partly reflects a selection bias that better students used the service 
more frequently.
Adjusting for this selection bias by using overall grade point average (GPA) as a classifier, we find a grade point improvement of $0.22$ for 
regular PR users, significantly above the random noise level of $0.04$.
Simply put, students who worked one or more additional problem per weekday 
earned nearly a quarter-letter average grade improvement irrespective of GPA.
Student comments emphasize the importance of randomness in helping them 
synthesize concepts.
The PR source code is publicly available on the {\em bitbucket} repository.

\end{abstract}

\maketitle 

\section{Introduction} 

Well-posed analytical problems and conceptual questions promote student growth 
from novice to expert learners \cite{Larkin80, Mestre93} and serve as the primary 
basis for evaluating student learning in physics and many other science and 
engineering disciplines.
In introductory courses, such evaluations usually take the form of timed examinations 
comprised of a number of independent problems.
Such examinations pose days of reckoning for students.
For an entire class period, perhaps longer, they must focus on correctly 
answering a set of questions that probe how well they understand and can apply the 
conceptual principles of the discipline they're attempting to master.

Good examinations explore interrelated concepts from a variety of angles.
For any particular problem, a student must be able to recognize which principles and 
methods are involved, execute a solution using the information 
provided, and do so in a timely manner.
For example, the first problem on a mechanics exam might ask the student to determine 
the value of a kinetic friction coefficient given an initial kinematic state and a sliding 
distance for some system.
Concepts and methods associated with forces and work are required, and a few lines 
of algebra then reveal the solution's form.
The next exam problem may address a completely different concept in a different way, 
perhaps by posing a qualitative question about angular momentum.
Given this type of evaluation environment, working old exam problems in random 
order seems a natural study strategy.

Introductory physics course enrollments at four-year colleges and universities have 
grown in recent years, reaching 452,000 in 2010-11\cite{aip2013report}.
At many universities, increasing student-to-faculty ratios precipitated a change in the 
nature of examinations, away from multi-part problems graded by hand and toward 
multiple-choice problems graded by machine.
The Physics Department at the University of Michigan (U-M) made this transition in the 
mid-1990's.  The past decade has also seen the rise of on-line learning services associated 
with popular textbooks, such as Pearson's Mastering Physics\cite{mastering}.  

The University of Michigan is a large, public university with undergraduate enrollment 
of 27,400 in the 2012-13 academic year \cite{umAlmanac}.
U-M Physics offers a pair of large-enrollment, two-term sequences, oriented toward Life 
Sciences and Science/Engineering students, respectively, that enroll roughly 1800 
students per term across the four courses.  Students in all courses currently use Mastering 
Physics for weekly homework problem sets.  
Each course also holds four examinations of $1.5$ to 2 hour duration and consisting of 20 
to 25  multiple-choice questions.
The questions are collectively written by the faculty members leading the class.
 
Over five years of this activity, a bank of roughly 800 pseudo-independent examination questions develops for each course.  By pseudo-independent, we acknowledge that problem re-use does occur from semester to semester, but rarely is an exact copy used.  Instead, a problem scenario that may involve a specific calculation one term may be rephrased as a conceptual question in a different term.  The physical system may be similar or even identical (think {\sl box on an inclined plane} or {\sl current-carrying wire} here), but the tenor of the question is new, and especially so to novice learners.  
  
In 2011, we realized that we could collect digital copies of these examinations and 
build a library to serve as a reference for students.
Instead of building a static reference for self-directed study --- a service which 
textbooks already provide --- we decided to build a front-end server that randomly 
selects among a topical set of problems.
Therein lies the core concept of the Problem Roulette service. 

The aim of Problem Roulette (PR) is to improve student learning by offering easy, 
random access to a large body of topical, locally-authored problems.
Prior to PR, a faculty member teaching a large introductory course would typically 
upload one or two old exams to serve as a study guide.
Students often sought out more practice exams, and a grey market for these 
materialized on campus.
Students associated with certain campus organizations, such as sororities, fraternities, 
honor societies, etc., had preferential access to a larger number of past exams, hence 
a potential degree of unfair advantage.
In this regard, PR serves as a social leveler, offering equal access to all students.

An advantage of the PR model noted by students is the random access mode of 
encountering problems.
While intentional, directed study of topics in the text or one's class notes is certainly 
beneficial, students who employ only this approach are potentially unprepared for 
problems that combine principles in ways not explicitly addressed by the these 
sources or problems that explore conceptual understanding in the context of a 
completely novel physical system.
Random access to problems outside of the textbook offers students a means to 
assess their level of comprehension.  More practically, it allows them to assess how  prepared they are for the upcoming exam.  

In a recent review of ten learning techniques popular in the educational psychology 
literature, {\em practice testing} and {\em distributed practice} (learning partitioned into 
multiple sessions) are the only two techniques rated as having high utility\cite
{Dunlosky13}.
Problem Roulette supports both of these techniques.
Students can effectively take a practice exam by working a series of questions within a 
time limit.
They can also break their study into chunks, distributing their learning across days and 
weeks in a convenient manner.
The PR service is also easy to use on mobile devices;  one student reported a habit of 
working problems using their cell phone while riding the intercampus bus.

In \S\ref{sec:architecture}, we present the architecture of the PR service, including its 
reporting capabilities.
We show usage patterns for Fall 2012 introductory courses in \S\ref{sec:analysis}, and 
examine the impact of regular use on final grades for the second semester Science/
Engineering course.
Potential future extensions are discussed in \S\ref{sec:future}, and we summarize in \S
\ref{sec:summary}.

\begin{figure}[h]
\centering
\includegraphics[width=0.8 \textwidth]{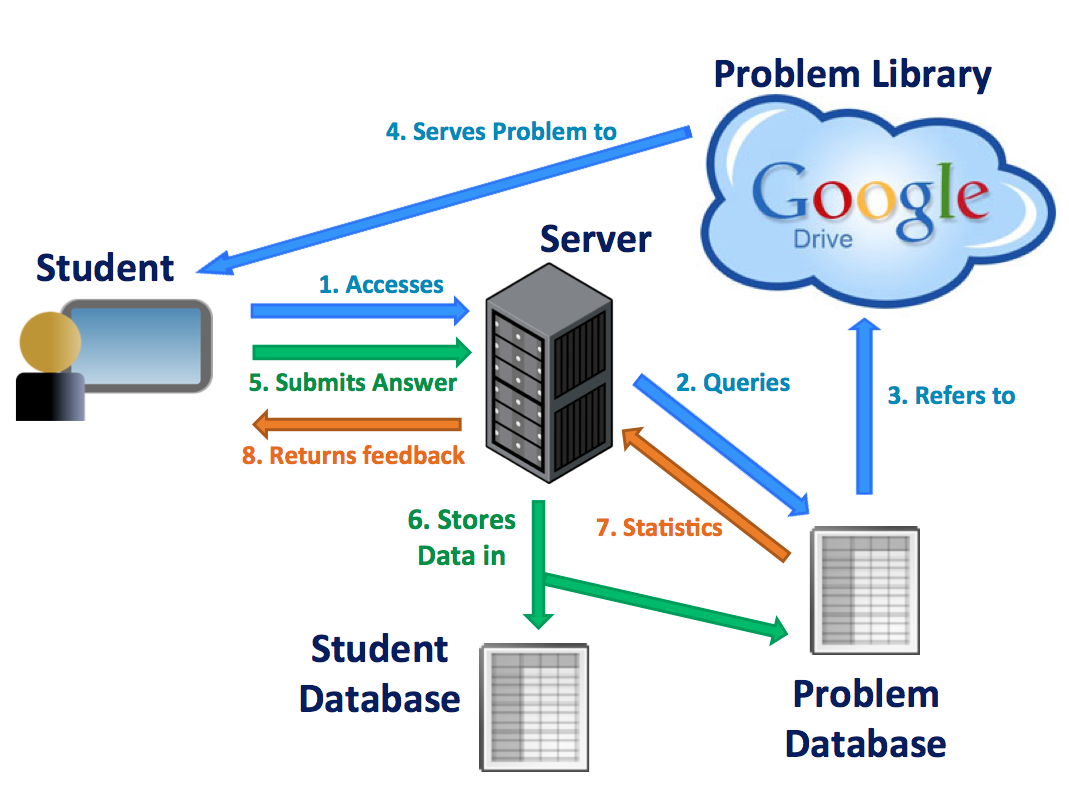}
\caption{Functional view of the PR service.  A student authenticates to a U-M server (1) 
and, after choosing a course/topic pair, receives a random problem pulled from the 
target library (2--4).  Student submission of an answer (5) triggers updates to the 
problem and student databases (6).  The server provides feedback to the student (7), 
an example of which is shown in Figure~\ref{fig:screenShots}.  }
\label{fig:PRservice}
\end{figure}

\section{Service Architecture} \label{sec:architecture}

Problem Roulette is a web service built atop hybrid (public-private) 
cloud computing infrastructure.
The basic content element is a single multiple-choice problem scraped from a previously 
administered examination and published as a public cloud asset in Google 
Drive. The collective set of problems forms a problem library organized by course, topic and 
other attributes.
A virtual web server in the private U-M cloud holds the set of problem locations 
(URL's) and answer keys.  
This server also manages student authentication using the universities 
Single Sign-On service and keeps the individual student transaction records.

Figure~\ref{fig:PRservice} illustrates how the service works.
After entering the site, a student chooses the particular course to study, then selects a 
topic within that course.
The topic structure, defined further below, can be general.
Our initial set-up is focused on exam preparation, so topics are listed by exam identity.
The student chooses among one of three midterms, a final, or all combined.
This selection then figuratively ``spins the roulette wheel'' to access one of typically 
hundreds of past, multiple-choice problems in their selected course/topic.
The server delivers a problem to the student's browser along with answer choices (for example A-E)
and the student can answer or spin again to select another problem. 

\begin{figure}[t]
\centering
\includegraphics[width=0.45 \textwidth]{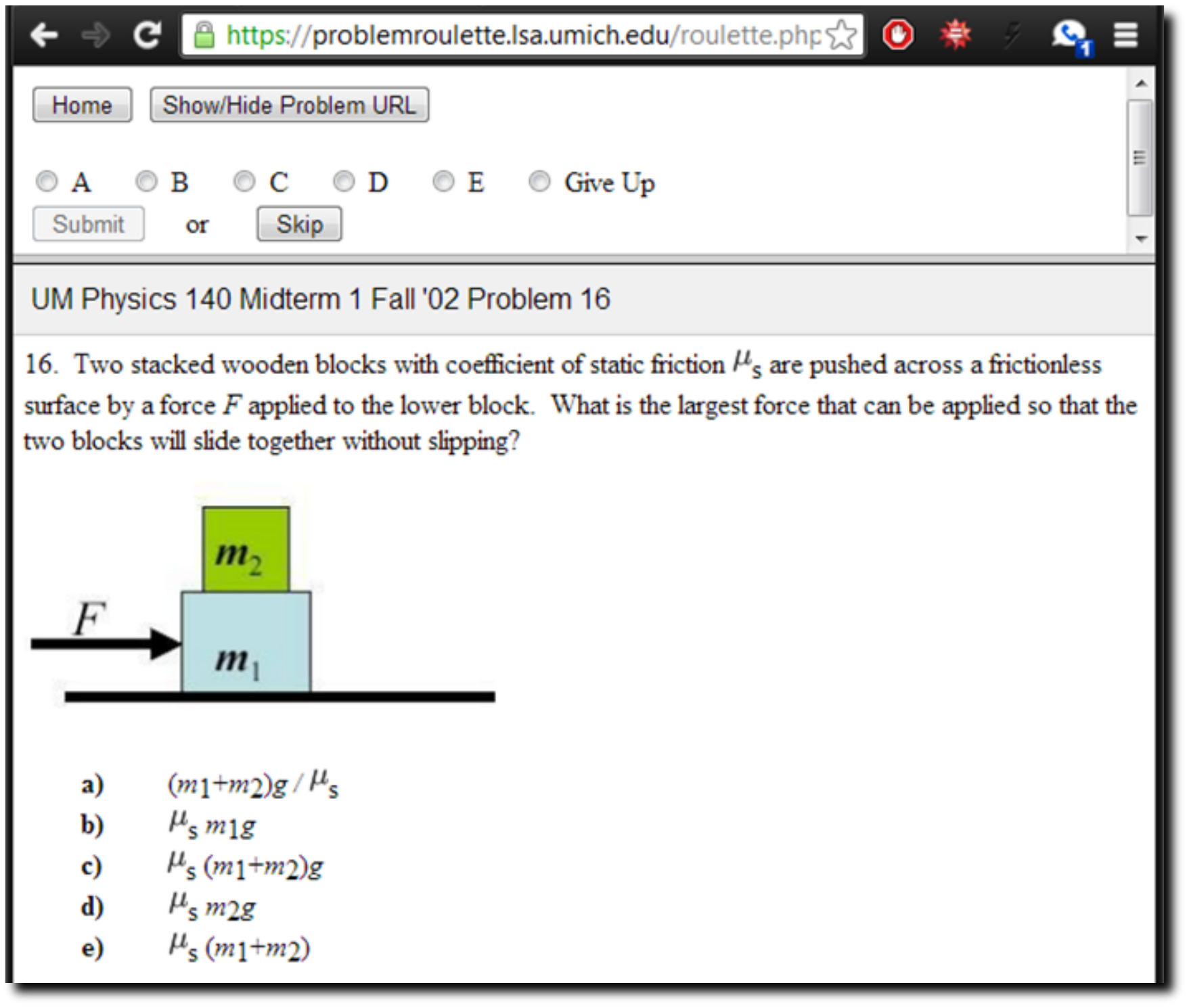}
\includegraphics[width=0.45 \textwidth]{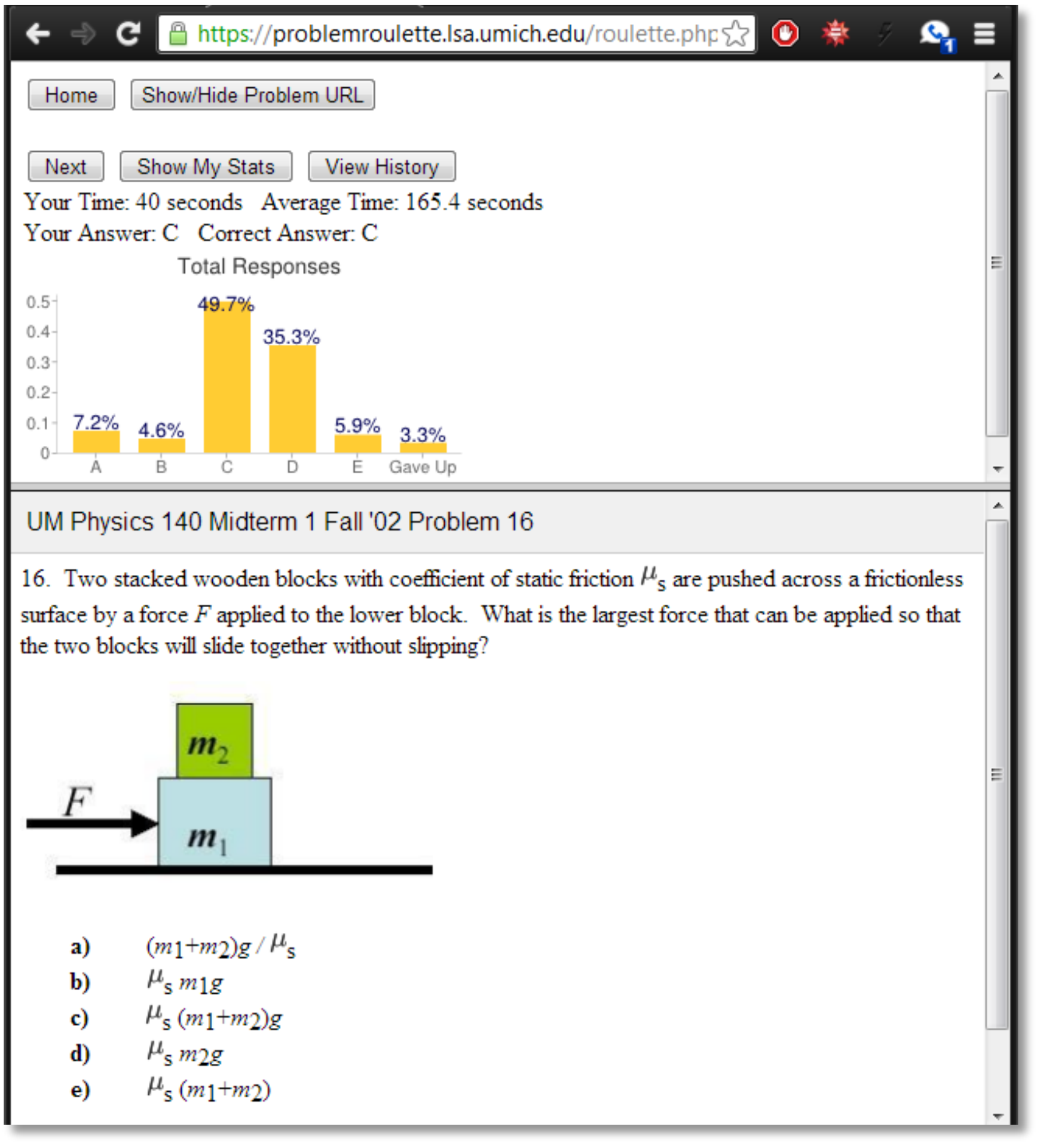}
\caption{
{\bf Left:}  Screenshot of a random problem from a Mechanics course.  The student can 
choose to submit an A--E answer, Give Up, or Skip.  A link to the problem URL is also 
available for future reference.  {\bf Right:}  Screenshot after a submission of  ``C'' 
includes  feedback to the student that: i) the answer of C is correct; ii) fewer than half of 
previous students answered correctly; iii) she/he answered comparatively quickly.   
Note that this example shows font formatting ambiguities introduced during the 
conversion from MS Word to Google Drive, visible in the answer list.  For example, 
answer {\bf b)} should read $\mu_s m_1 g$. }
\label{fig:screenShots}
\end{figure}

Figure~\ref{fig:screenShots} shows screen shots of an example problem as 
served (left panel) and after the student responds with the answer of ``C'' (right panel).
The student's selection is recorded in a transaction database, and she/he receives 
immediate feedback beginning with whether the answer is correct (in this case, 
C is indeed the correct answer).
The student also receives a frequency histogram of peer-submitted 
answers.  For this example the plurality, but not majority, answered correctly.
Using the time interval between serving the problem and receiving a response as the solution time, the mean time to solution for peers is also returned for comparison to the student's own time.  
For the example shown, the 40 second answer was considerably faster than the 
average of 165 seconds
In summary, this student got a problem correct that most students did not, and did it 
faster than her/his peers.

A problem \textit{session} consists of the set of $N_p$ problems worked in succession 
by a student before logging out or being timed out after an hour of inactivity.
Session lengths vary, the frequency distribution scales approximately as $N_p^{-1}$ 
and falls more sharply for $N_p > 30$.
One enterprising student worked 92 problems in a session lasting two hours and 
eleven minutes, a brisk clip averaging 85 seconds per problem.

Within a single session, the random selection of problems is done without 
replacement.
However, each session begins anew using the full problem stack, so students who 
return to study multiple times are likely to repeat problems.
We show below 
how the correct response rate rises and 
response time falls as the same problem is encountered across multiple sessions. 

\subsection{Problem Library} 

We reached out to U-M Physics faculty and received dozens of old exams in either 
MSWord or pdf format.
Each problem on the exam is scraped and uploaded as a unique document on a 
cloud storage service.
We use Google Drive (formerly Docs) as the storage solution, but any service capable 
of serving documents as web pages could be used.

We hired advanced undergraduate students to upload problems manually, as there 
are formatting issues associated with super- or sub-scripts, graphics, and greek 
symbols that do not always transfer correctly (see Figure~\ref{fig:screenShots}).
Average upload time is roughly 5 minutes per problem, so a 20-question exam can be 
completed in under two hours and five semesters of exams (400 problems) can be 
converted with roughly one week of effort.
Overall, we have converted more than 1200 problems and we plan to continue adding 
new exam problems at the end of each academic year.

A clean view of the problem in a web browser, devoid of the toolbar and other editing 
options available to the owner, requires that the document be formally published on 
Google Drive\cite{googlePublishHelp}.
The action of publishing creates a random-string URL for each problem, which is 
captured and added to the server database.
Currently, file ownership is shared among PR developers, who have authority to 
correct errors or otherwise modify content.


\subsection{Server Configurations} 

PR is deployed in a LAMP (Linux, Apache, My-SQL, PHP) environment 
on a virtual host provided by Informational Technology Services (ITS) at the 
University of Michigan \cite{umWebServices}.

The lightweight PHP application simply manages the web transactions and does not require any 
special framework.
We provide more detail and reference to the open source code in the Appendix.
A standard server configuration for such a LAMP system is readily available on 
Amazon Web Services \cite{AWSEC2}, so a PR application could be hosted in the 
public cloud for a modest fee.
Our private cloud version hosted by ITS simplifies authentication for U-M students 
at a modest cost to the Physics Department of \$50/year. 

The PHP application follows an object-oriented Model-View-Control (MVC) design pattern. 
With an MVC design, the code is partitioned into three different parts designated for models, views, and controls.
For example, model classes represent objects such as problems and users so that they have the appropriate 
properties and methods to be easily managed in the PR application.
Views classes represent visible elements and deliver display components which in this case are blocks of html, css, and javascript to create the web pages.  Control handlers are 
the php files that handle server requests, generated for example by clicking to select a particular 
topic or submitting an answer.  Controls manipulate the model and view objects to maintain 
application state and produce updates to the interface.  This design pattern structures the 
code to make it more readable, reusable, and robust.
 
\begin{figure}[t]
\centering
\includegraphics[width=0.75\textwidth]{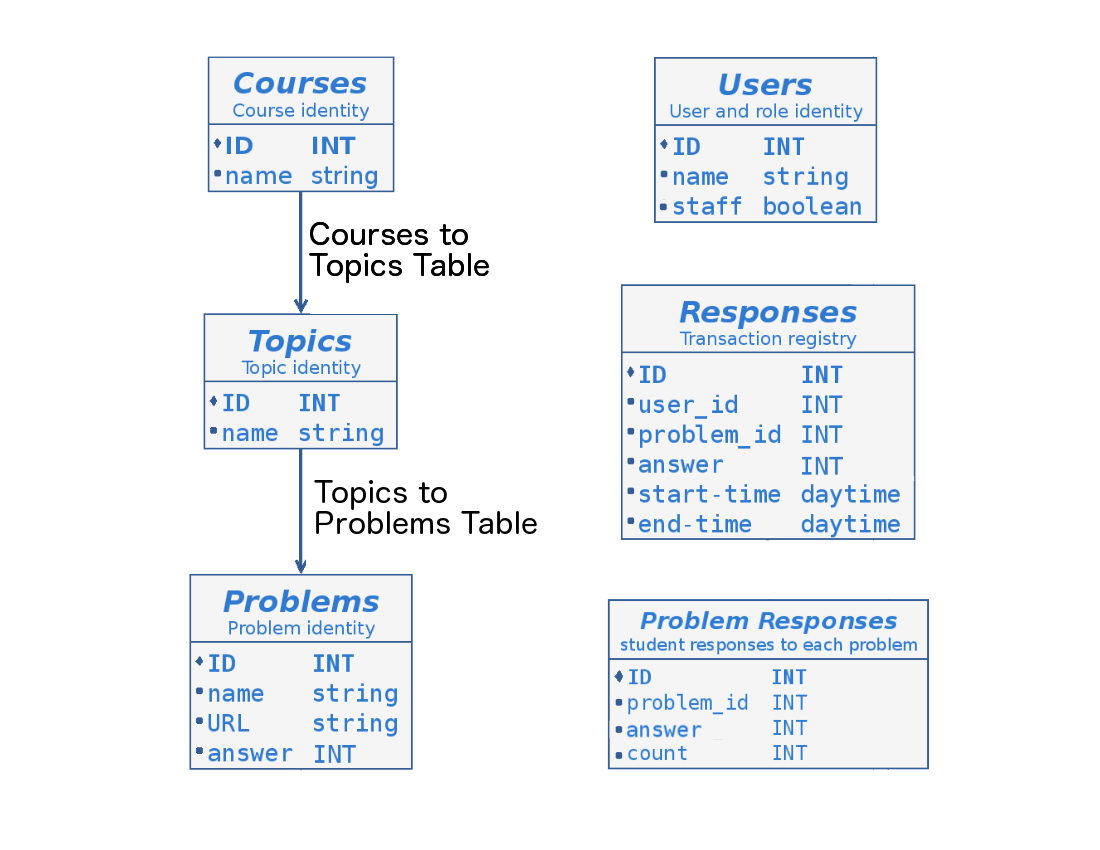}
\vspace{-14.4truept}
\caption{Illustration of the PR database schema shows the main data tables and relationships, 
with mostly static elements in the left column and  dynamic elements at right.  The 
``Courses to Topics Table'' table has just two columns and provides a mapping of course ID's to topic ID's.
Similarily the ``Topics to Problems Table'' provides a mapping from topics to problems.  Unique topic ID's will generally 
map to one course while individual problems may map to more than one topic.  The 
Responses table is the core transaction table that records student activity.  The history 
of responses to each problem are updated dynamically and the summary data about responses are cached in the Problem 
Responses table for fast, real-time access.}
\label{fig:PRschema}
\end{figure}

The database tables and schema in Figure~\ref{fig:PRschema} serve to elucidate the MVC model so we describe them here.
Database tables contain lists of Courses, Topics, and Problems.
Course names follow the U-M course catalog.
Topics are currently named by exam (Midterm 1, 2, 3 or Final) but the topic list can be 
readily extended to cover particular physics concepts, chapters in a textbook, or other 
classifications.
A relational table links the topic entries to corresponding courses.

The Problems table holds the name, URL, and correct answer of a problem.
The Problem Responses table stores the response choices for each problem and dynamically updated
cache values for how many times each response was submitted by students.
The latter is used to create the frequency distribution graphic reported back after an 
answer is submitted (right panel in Figure~\ref{fig:screenShots}).
A relational table links problem entries to corresponding topics such that a given problem can map 
to multiple topics.  

The Users table holds login names of students and staff, with the latter having 
privileges to add and modify content and to extract student and problem records.  The 
service is not limited to enrolled students; any member of the university community with 
valid authentication credentials can use it.  
 
\subsection{Response Table, History Tool, and Analytics} 

At the core of the service lies the transaction table of student interactions.
When a student answers a problem, the Responses table is updated to record the 
User ID, problem ID, student response and timestamps recording when the problem 
was served and when the response was submitted.
This general structure allows rich, post hoc data analysis, some of which is described below. 

A history view is currently provided to students which shows them a sortable list
of past problems and performance.
Students find the ability to reference problems they've answered incorrectly a 
convenient feature during visits to office hours.
The fact that this record is stored on the server makes it is readily available at any 
time from any networked device, including a cell phone.  Students can consult with their peers on problems easily, without the need to carry around a heavy, marked-up textbook.

For analysis of student performance, we have access to additional information from 
our course management system as well as from E$^2$-Coach \cite{E2coachSite}, a 
new student coaching system being developed at U-M as part of a Provost's learning 
analytics initiative.
Along with exam scores and final course grade, E$^2$-Coach holds additional 
student record information, such as pre-college measures, demographic information, 
and overall grade-point average (GPA) at Michigan at the beginning of term.
We use the latter to assess the influence of selection bias on the grade performance of 
regular PR users.

\section{Problem Roulette Usage and Impact } \label{sec:analysis}

\begin{table}[t]
\centering
\caption{Introductory Course Enrollments and PR Usage in Fall 2012}
\begin{tabular}{l c c c }
\\
\hline
Course (Name) & $N_{\rm enroll}$ & $N_{\rm PR1}$  & $N_{\rm PR50}$ \\
\hline
\hline	
Science/Engineering, 1$^{\rm st}$ semester  (Phys 140) & 571 & 367 & 97  \\
Science/Engineering, 2$^{\rm nd}$ semester  (Phys 240)  & 531 & 456 & 229  \\
Life Sciences, 1$^{\rm st}$ semester (Phys 135)  & 474 & 124 & 34  \\
Life Sciences, 2$^{\rm nd}$ semester  (Phys 235)  & 263  & 23 & 0    \\
\hline
\end{tabular}
\label{tab:introCourses}
\end{table}

A beta version of PR --- one without the ability to capture student responses --- was 
released as an optional study service in the 2011-12 Science/Engineering sequence 
and proved to be popular.
The redesigned (v1.0) service described here was expanded in Fall 2012 to serve 
both the Science/Engineering and Life Sciences sequences in the Physics 
Department.
 
Table~\ref{tab:introCourses} lists the courses and their overall enrollments.
The quantities $N_{\rm PR1}$ and $N_{\rm PR50}$ are the number of enrolled 
students who answered at least one or fifty or more PR problems, respectively.
The level of engagement is larger in the Science/Engineering courses mainly because 
of the larger problem libraries available. Peer publicity from previous and current 
students also plays a role;  Google Analytics traffic shows students arriving via links 
shared on Facebook.
 
In the analysis below, we use the entire sample of all four classes except when we 
evaluate the impact on student grade performance.
To study grade outcomes, we focus on Phys 240 because it has a larger problem library and because most students in that class have meaningful beginning-of-term U-M GPA values.

\subsection{Service Activity in Fall 2012}

\begin{figure}[h]
\centering
\includegraphics[width=0.95 \textwidth]{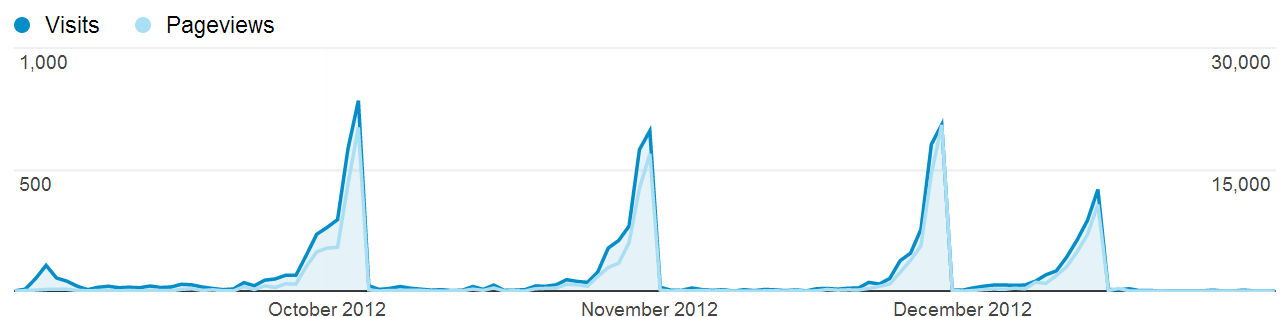}
\caption{Number of PR visitors (dark line, left axis) and page views (light line, right 
axis) per day for the Fall 2012 term.  Activity clearly spikes on the days leading up to 
the four examinations.   
}
\label{fig:analytics}
\end{figure}

We stress that there is no formal course credit associated with use of PR; it is an 
optional study aide only.
We are encouraged, then, by the fact that the majority of students in the Science and 
Engineering course sequence (Phys 140/240) used the service at least once, and that 
more than 25\% used it on a regular basis ($\ge 50$ problems over the term).
We share below some student narratives that explain their views of the utility of PR.

Overall, we served over 60,000 pages of problems to nearly 1000 students during the 
four month period, September--December 2012.
Figure~\ref{fig:analytics} shows that PR usage by students varied considerably over 
the course of the term, with spikes in activity occurring immediately prior to the three 
midterm examinations and the final exam.
The sharpness of the peaks is enhanced by the fact that, for room scheduling reasons, 
all four courses hold evening exams on the same set of dates.

\begin{figure}[h]
\centering
\includegraphics[width=0.5 \textwidth]{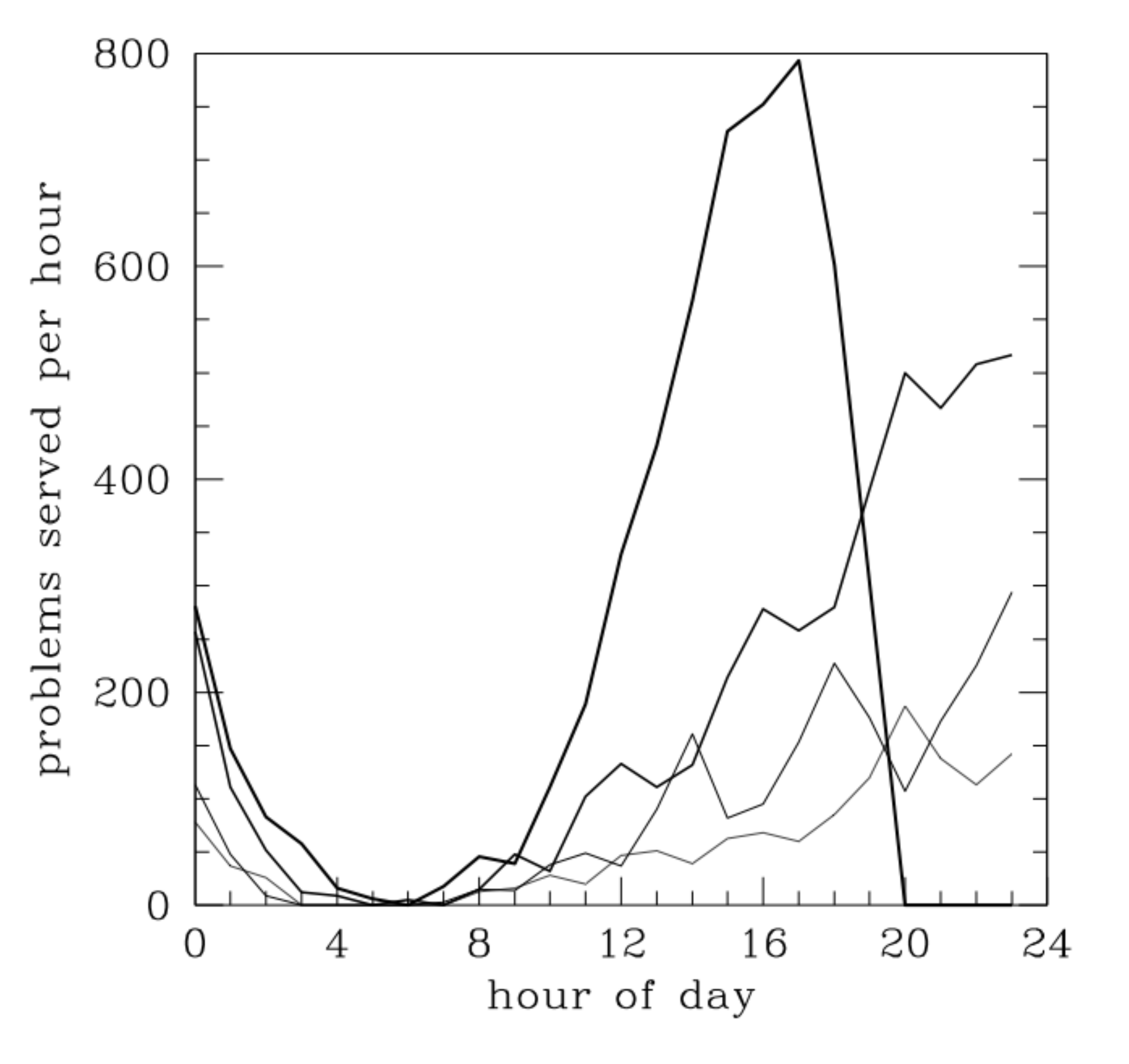}
\vspace{-14.4truept}
\caption{PR activity on the Monday (lightest line), Tuesday, Wednesday and Thursday 
(boldest line) of the first midterm week.    Examinations start at either 6:00 or 8:00 pm 
on Thursday. 
}
\label{fig:examWeek}
\end{figure}

In Figure~\ref{fig:examWeek} we show the buildup of student engagement over the 
four days preceding the first midterm examination of the term.
The volume of activity in the 24 hours preceding the exams nearly equals the activity 
of the three days prior.
From this evidence, we may surmise that students consider PR as more of a {\em 
cramming} service than a study service.
But anecdotal evidence suggests that students working problems the day of the exam 
include a large fraction of regular users sharpening their skills.
Like an athlete preparing for a game, they are warming up with the aim to hit the exam 
room already in the groove.


Given the topical design of arranging problems by examination, the usage 
patterns seen in Figures~\ref{fig:analytics} and \ref{fig:examWeek} are hardly surprising.
Re-alignment of the service along physics-oriented topics might be a way to promote 
more regular study throughout the term.

\subsection{Student Behaviors and Perceptions} 

\begin{figure}[t]
\centering
\includegraphics[width=0.48 \textwidth]{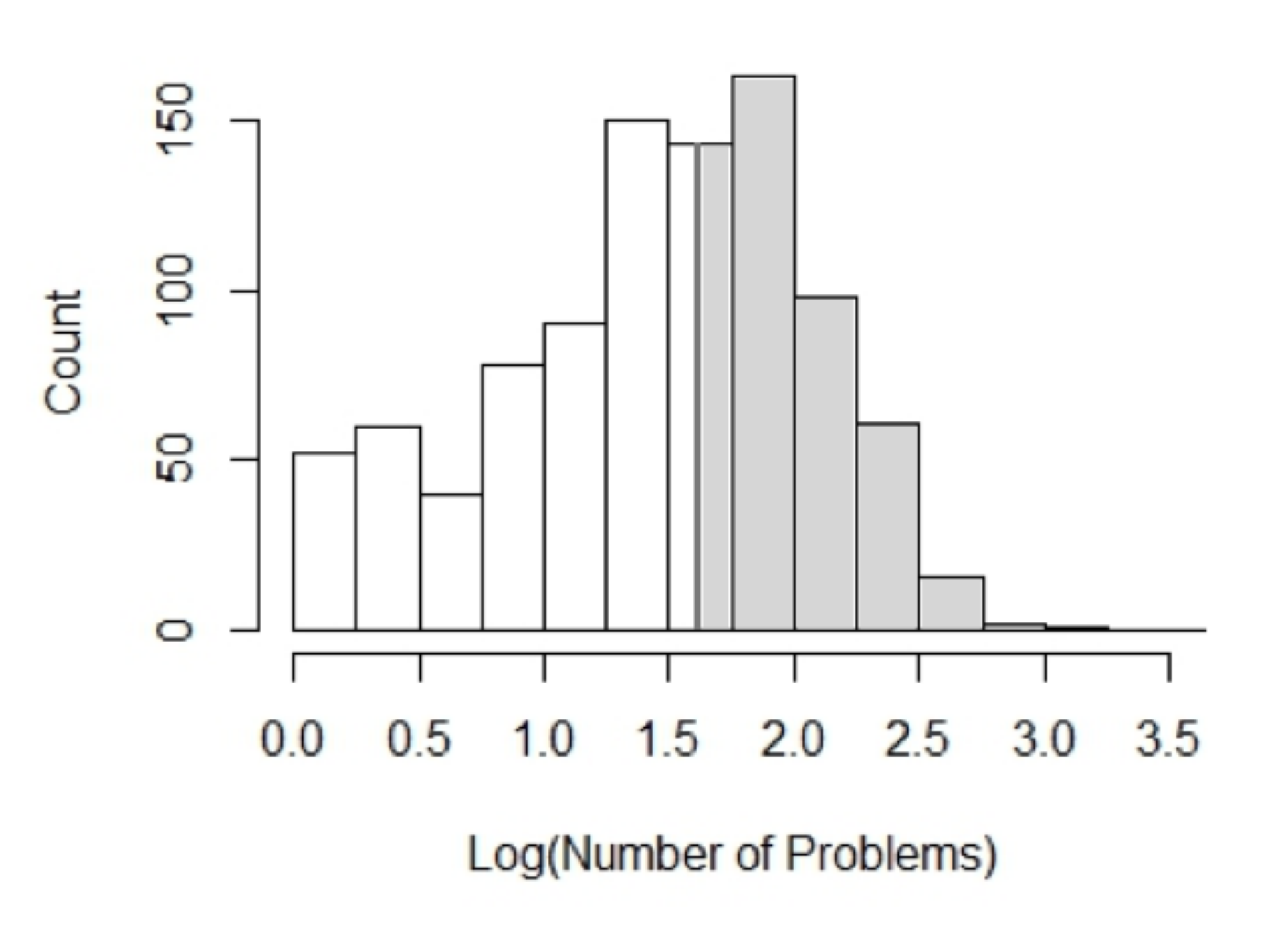}
\includegraphics[width=0.48 \textwidth]{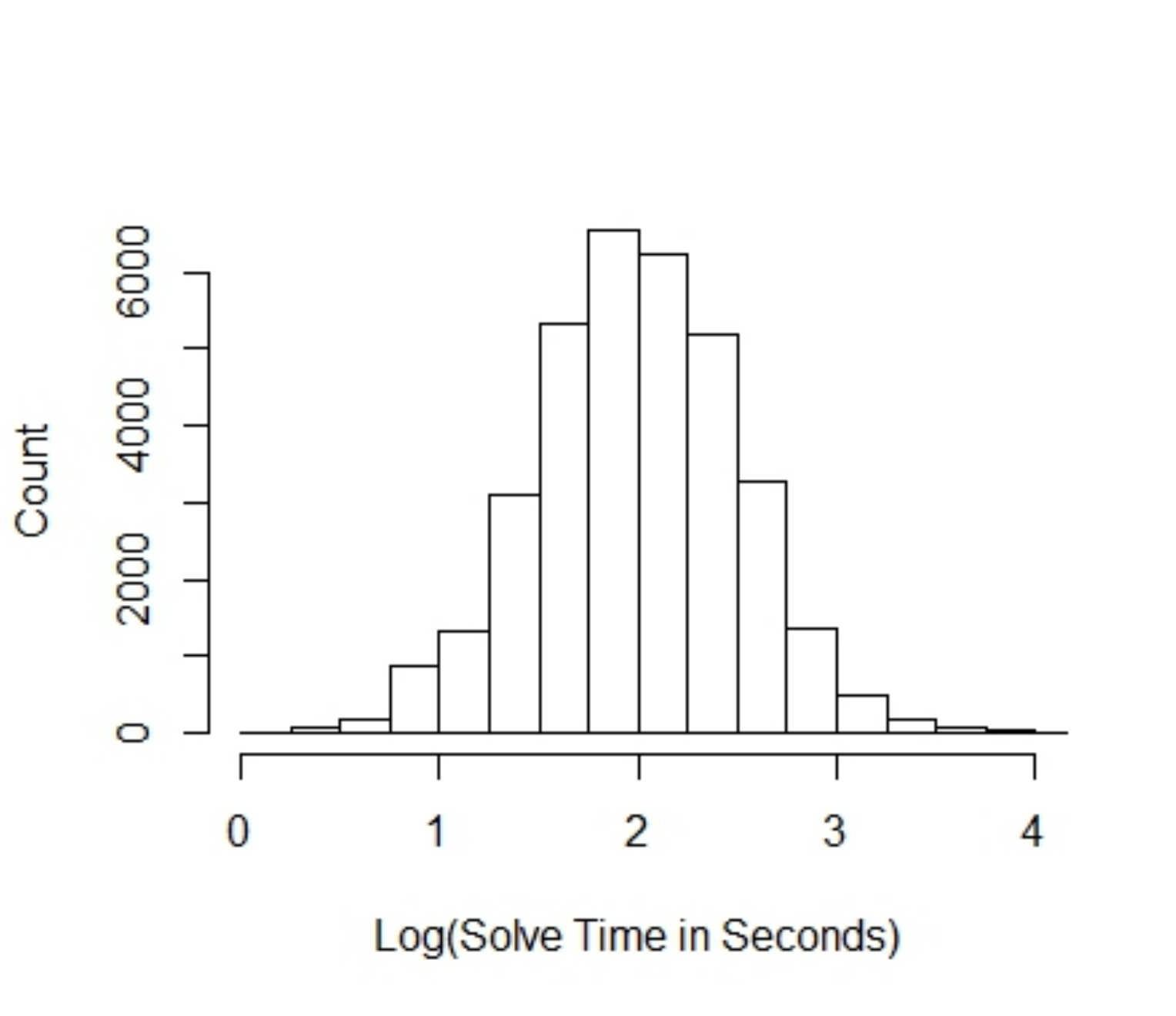}
\caption{{\bf Left:}  Distribution of the overall number of problems worked by individual 
students during Fall 2012.  Note the decimal logarithmic scale.   Students who work 
more than the median value of 50 problems are considered  {\em regular\/} users 
(shaded area).  Twenty students solved over three hundred problems during the term.  
{\bf Right:}  Distribution of response times for all problems worked by students in Fall 
2012.  The median value is close to 100 seconds.  
}
\label{fig:basicStats}
\end{figure}

Figure~\ref{fig:basicStats} provides views of statistics on overall PR usage by students 
and on the solution time per problem.
The left panel shows the frequency of students characterized by the total number of 
problems, $N_t$, each one worked during the term.
The distribution in logarithmic $N_t$ bins gently rises up to a few hundred problems 
beyond which it falls off.
One enterprising student worked over 1000 problems (including repeat occurrences) 
during the term.
We use the median value of 50 problems as a minimum value to define a set of {\sl 
regular} users of PR.
We discuss this cohort in more detail below. 


The right panel of Figure~\ref{fig:basicStats}  shows the frequency of response times, 
again using logarithmic binning in time.
The median response time is quite short, approximately 90 seconds, and the 
distribution is approximately log-normal with dispersion in decimal log of $\sim \! 0.4$.
The relatively short median response time reflects several factors.
A significant fraction of the multiple-choice questions are conceptual, and many of 
these can be read and answered very quickly.
For example, the question with the shortest mean response time, and not 
coincidentally also a very high correct response percentage, is the following --- 
``Newton's first law of motion is most closely associated with what fundamental 
physical property of matter?" --- with answers:  A) inertia; B) angular momentum; C) 
weight; D) density; E) rigidity.
This problem can be read and understood in a matter of seconds.

A second factor bringing down the median time per problem is repetition.
Students may see the same problem in different sessions, and familiarity leads to 
decreasing response times, as  shown below.
A third possible factor is that some students may submit a random answer quickly in 
order to reveal the correct answer which appears after their submission.
The lack of a spike at very short response times, however, implies that this is not a 
widespread mode of behavior.

Rather than repetition, which does occur at a low level, it is the random access nature 
of the PR service that students find valuable.
Students attest that working a set of randomly-chosen problems helps them learn and 
synthesize concepts, and they realize that this is critical for their success in physics.
The following pair of comments are illustrative. 

\begin{quote}
{\sl 
Before the first midterm, I just relied on the practice exam and problem sets I had 
completed previously; when I was looking over my old  problem sets, I recognized that 
I was doing them partially through rote memory, as I had done them all before. I did a 
bit above average on the first midterm, so I decided to give PR a whirl for extra practice.  
The refreshing thing about PR was that I could actually practice all the formulas and 
concepts I had learned in class on new problems; no longer relying on memorization, 
the concepts actually stuck with me at exam time, and I ended up getting 95 and 100 
and my next two midterms.}
\end{quote}

\begin{quote}
{\sl 
For the first exam I looked at problem roulette and only understood around half of the 
problems but I figured it was just because they were the hardest of th{\rm [eir]} type. 
I did not do so well on that exam and for all the future exams I focused more on 
problem roulette and once I understood how to answer all of those questions I 
understood all the concepts and thus all the exam questions.
}
\end{quote}

The reporting back of problem statistics after submission by the student is also a 
popular aspect, as exhibited by these two student testimonies.

\begin{quote}
{\sl 
I really enjoyed using the Problem Roulette to study because it allowed me to do 
practice problems and compare my skills to others'. 
It was interesting to see where other people took more or less time than I did, on each 
specific problem. 
Also, it was useful to see the histogram of the answers, to see if my mistake was 
common or not. 
}
\end{quote}

\begin{quote}
{\sl 
Among the benefits of problem roulette is it shows you statistics of how other students 
answered the question and how fast they did so. 
This lets you know if you are above average or below average on topics and what you 
need to study.
}
\end{quote}

\subsection{Impact on Course Grade} 

How does regular PR usage influence course grade performance?   
To address this question, we restrict attention to the 2nd semester Science/
Engineering course, Phys 240 because the majority of students in this class have a 
non-zero U-M GPA at the start of term.   
We use this measure as a classifier to adjust for the selection bias associated with 
stronger students using the service more.  

\begin{figure}[t]
\centering
\includegraphics[width=0.6 \textwidth]{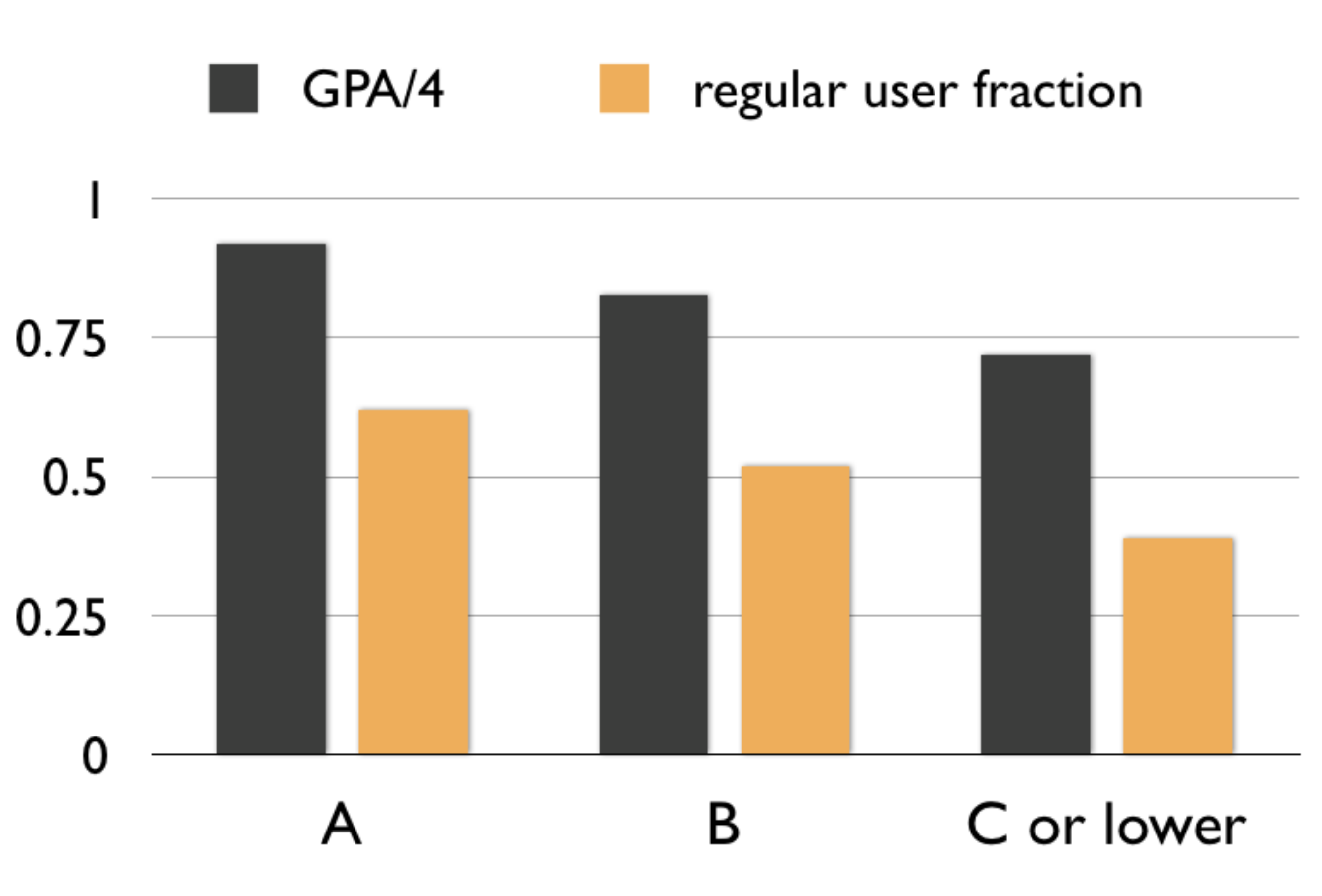}
\vspace{-14.4truept}
\caption{This figure illustrates the selection bias of regular PR users in the Fall 2012 
electromagnetism course (Phys 240).  
For students who achieved the final letter grades shown, we plot both the average, 
overall Michigan GPA at the beginning of Fall term (divided by 4.0, black) and the 
fraction of regular PR users (light) within each letter grade population.   
Higher-performing students in Phys 240 tend to be higher performing overall (higher 
overall GPA) and also tend to use the PR service more.   
}
\label{fig:GPAbias}
\end{figure}

Splitting the class enrollment into regular PR users (229 students) and the remainder 
(289), we find that regular users outperform their complement by an average of 0.40 
grade points, nearly half a letter grade.  
But this analysis is overly simplistic, as it does not address the fact that the optional 
nature of the PR service means that students are free to select their own degree of 
engagement.  

Figure~\ref{fig:GPAbias} demonstrates that there is a selection bias.  For students earning final letter grades of A, B, or C or lower, we show both the mean  
beginning-of-term GPA (divided by 4.0 to provide a percentage) and the fraction of regular PR users within each letter grade subset.  Of the 518 graded students in the Phys 240, 156 received A's, 168 received B's, and 194 received C's or lower.   
There are clear trends with letter grade.  
The regular user fraction of A students is $62\%$, declining to $39\%$ of C or lower 
students.  
At the same time, the mean GPA drops from $3.67$ to $2.88$ across those categories.   

Considering overall, beginning-of-term GPA as a proxy for student ability, the 
evidence indicates that the cohort of regular PR users is disproportionately comprised 
of higher-ability students.  
We therefore need to factor this selection bias into analysis of the impact of regular PR 
use.  

\begin{figure}[t]
\centering
\includegraphics[width=0.6 \textwidth]{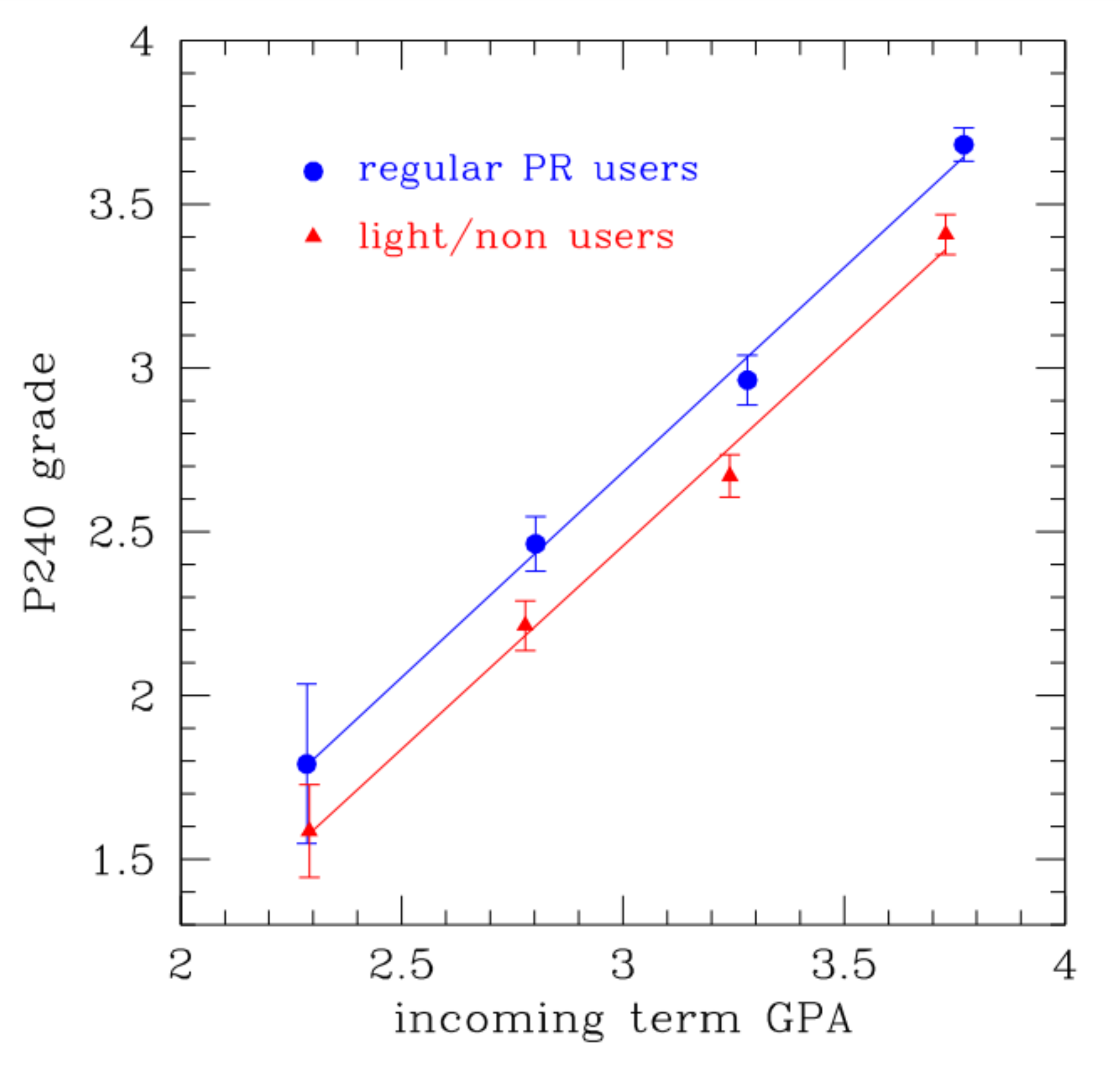}
\vspace{-14.4truept}
\caption{
Grade points earned in Physics 240 are shown separately for regular (blue circles) 
and for light/non (red triangles) users of the PR service.  
Students are partitioned into four bins based on beginning of term GPA.  
Points are plotted at the mean GPA value of each subset, and error bars show the 
standard deviation of the mean grade.  
Linear fits to the mean measures are shown.  
Regular PR users are offset at a level of approximately 0.23 in final grade relative to 
light/non users.   
}
\label{fig:examBoost}
\end{figure}

To mitigate this selection bias, we use beginning-of-term GPA as a classifier.   
In Figure~\ref{fig:examBoost}, we display the final grades of students divided into 
different beginning-of-term GPA bins and sub-divided as regular users (blue circles) or 
their complement (red triangles).   
Points are plotted at the mean GPA and course grade within each sub-population, and 
error bars show the standard deviation of the mean grade.  
Lines are linear fits to the points shown, both of which have a slope of approximately 
5/4.  
(Independent analysis of this trend using larger samples indicates a degree of positive 
curvature, with low GPA grades lying somewhat higher than the linear trend indicates.)

Students of all abilities benefit from regular PR usage.  
The mean offset of the linear fits is 0.22 grade points, a boost of approximately a 
$3\sigma$ in all but the lowest GPA bin for which statistics are poor.  

In addition, we perform a null test where we draw random sets of students of size 
equal to the regular PR cohort, 229 out of the 518 graded students.  
Generating 10,000 random realizations, we find a mean grade point of $2.88$ and 
dispersion of $0.04$.  
The mean grade point of the regular PR users is $3.10$, displaced $5\sigma$ upward 
from the random mean.  
We repeated this exercise for Phys 240 in Winter term 2013 and found an effect of 
somewhat larger magnitude, $0.32$ grade points.  
The smaller class size of Winter term had a larger noise level of $0.06$, leading again to a $5\sigma
$ effect.  

Finally, we have performed a GPA-matching analysis where we sample the non-regular user cohort of students in a manner weighted to match the GPA 
distribution of the regular user sample.  
We again find a boost of 0.22 grade points for regular users, consistent with the result 
shown in Figure~\ref{fig:examBoost}.  


\section{Future Directions}\label{sec:future}

While the long-term evolution of home-grown services like Problem Roulette can be 
difficult to predict in a university environment, we discuss here ideas for progress in 
the near term.  

\subsection{Expansion and Extensions} 

Educational psychology studies highlight the strong value of practice tests and 
distributed study to promote student learning\cite{Dunlosky13}.  
The PR service incorporates both of these elements in a lightweight framework that 
supports study using mobile devices.   
Any course which administers tests in a choose-from-a-list format can establish a 
problem library within a topical structure and join the existing service simply by adding 
the relevant data elements to the server tables (see Fig.~\ref{fig:PRschema}).  
 
We are currently working with faculty in four additional U-M departments --- Chemistry, 
Biochemistry, Statistics and Atmospheric, Oceanic and Space Sciences  ---- to  roll out 
PR to their introductory classes in the 2013-14 academic year.   
A single server instance will support all the classes under a common interface and 
activity will be recorded in a single Responses table.   
The expansion is requiring us to refine staff roles and extend reporting functions.   

Potential extensions to the service are many.  
One idea is to make PR more social by allowing students to tag or rate problems.  
Tagging could be based on physics concepts, and any given problem would generally 
have multiple tags.  
The action of tagging a problem would force students to map problems to concepts, 
and the collective set of tags would produce a body of student associations that reflect 
their conceptual understanding.
Choosing and managing the taxonomy of allowed tags is a challenge for this potential 
extension, as current taxonomies designed for use by physics education researchers\cite
{Toedorescu13} are too high level for novice learners.  

Ratings are a potentially simpler extension.
On the page that follows submission of an answer, a student could be offered the 
opportunity to rate the problem.
Ratings could be on a Likert scale (1-5) and be based on the problem's complexity 
(1=Easy, 5=Hard), its perceived utility to learning  (1=Useless, 5=Helpful) or its clarity  
(1=Confusing, 5=Clear).

The ability to see fully worked solutions to problems is an obvious addition.
We have so far deliberately omitted this element in PR because our intent is to mimic 
the examination environment in which the student must work in isolation to produce a 
solution.
Technically, links to worked solutions can be added as a new item in the Problems 
table.
The Statistics department already has video solutions for prior examination problems, 
and we are working to incorporate them as part of the ongoing expansion.

While adding some or all of the above features may be attractive, it is important to 
recognize that the lightweight nature of the PR service has strong appeal.
Currently students can navigate into the service with three mouse clicks, and only two 
clicks per answered problem are required.
Extensions should be done in a way that respects this minimalist design philosophy.

\subsection{Education Research Opportunities}
 	
While the simple message that working more problems leads to improved learning  
should come as no surprise to most physicists and physics teachers, 
isolating the role that regular problem practice has in learning physics is a complex 
task.
The grade boost for regular PR users might, in part, be due to other correlated 
attributes that affect grade performance.

We see some evidence of this complexity in the Phys 240 data of Fall 2012.
Regular users outperformed their complement on exam scores, but only 
at a modest level of $3\%$.  
The effect on overall grade is thus only partly due to improved exam performance.  Higher scores 
in the other graded components of the course -- on-line homework and in-lecture 
response scores -- are also important.

One might conclude that regular users of PR also tend to be more conscientious about 
getting their homework done and participating in lecture.
However, limited sample statistics may also play a role.  The Fall 
2012 pattern is not as evident in the Winter 13 class of Phys 240, where we find that 
regular users score an average of 5\% higher on examinations.
More study on larger statistical samples is clearly needed to sort out 
interrelated effects and tease out causal relationships.  

\begin{figure}[t]
\centering
\includegraphics[width=0.31 \textwidth]{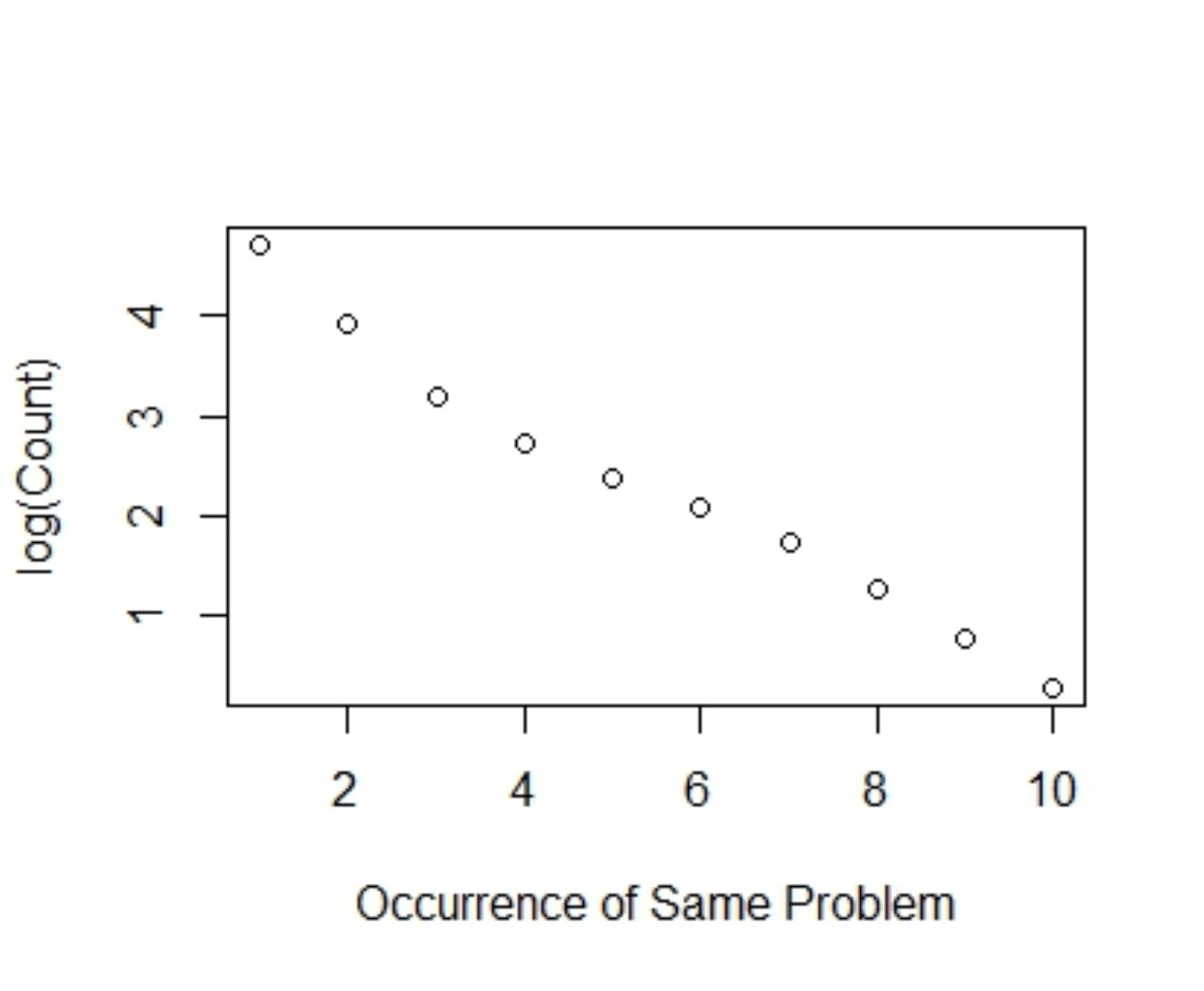}
\includegraphics[width=0.31 \textwidth]{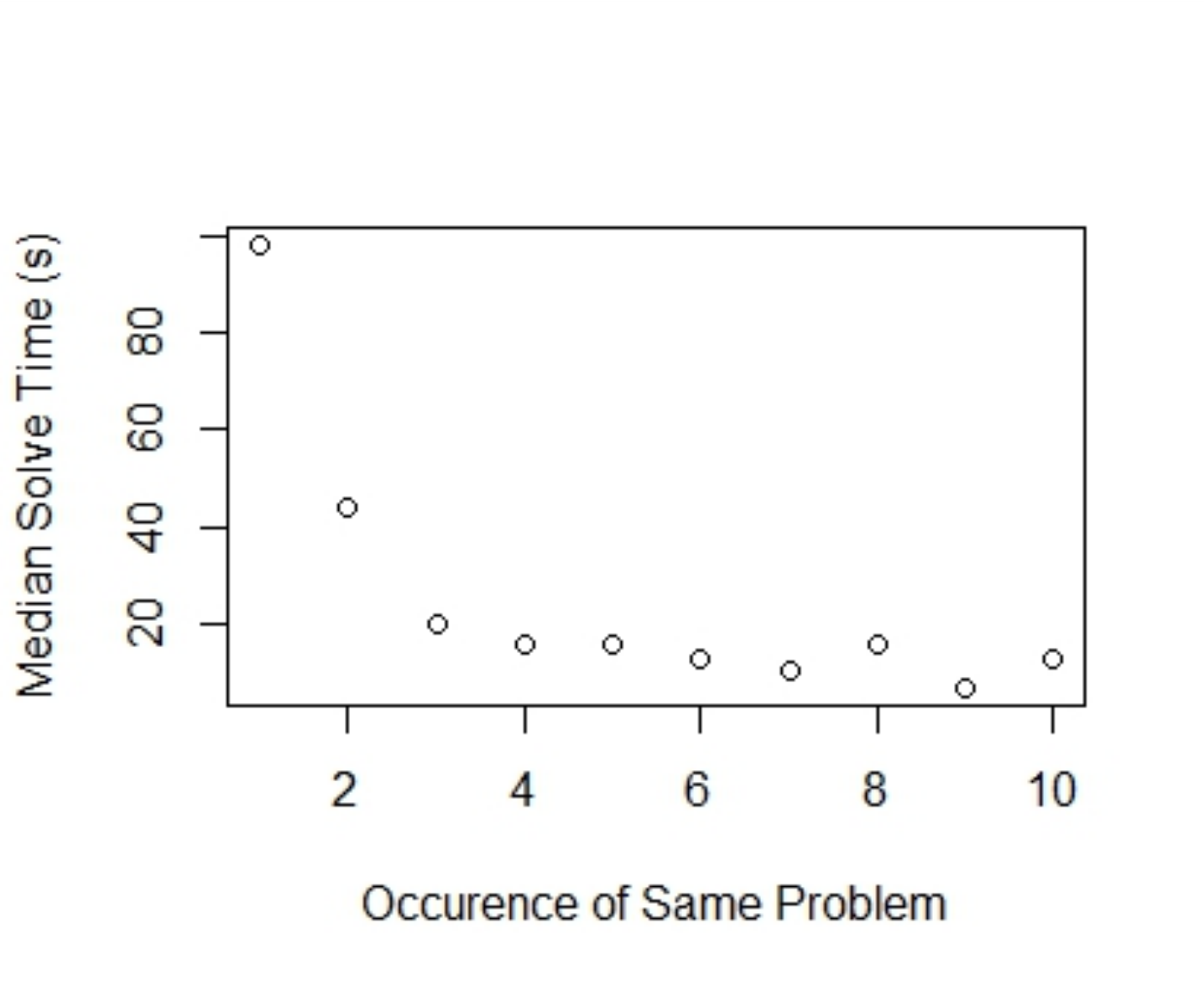}
\includegraphics[width=0.31 \textwidth]{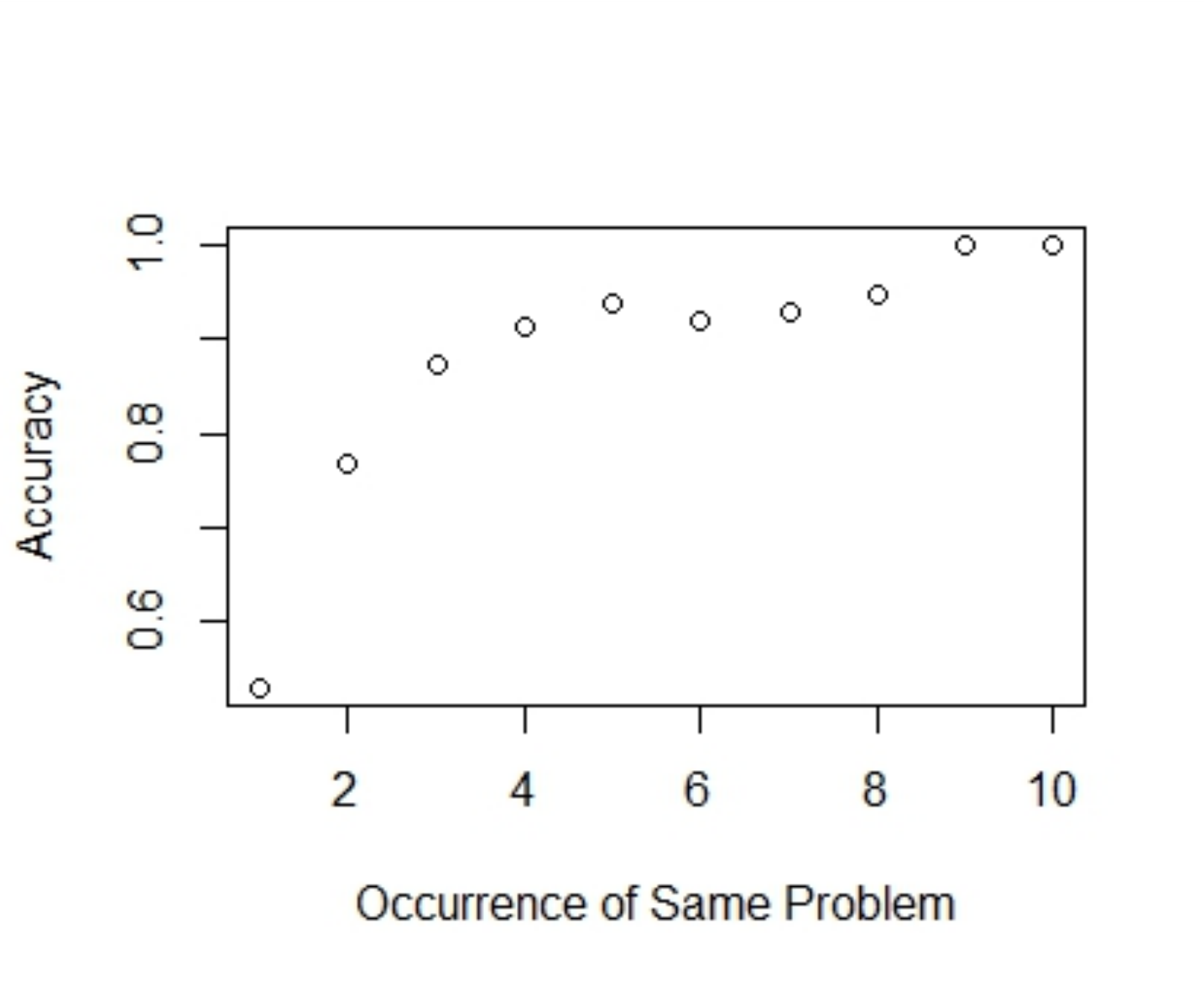}
\vspace{-14.4truept}
\caption{
Statistics derived from the Responses table can be used to address the issue of 
retention in problem solving.   
The left panel shows the number of times that students encountered a problem 
multiple times.   
The median time for solution (center panel) declines and the fraction of correct 
responses (right panel) rises as the same problem is encountered, with the largest 
gains concentrated in the first few occurrences.  
}
\label{fig:repeatProbs}
\end{figure}

The collective set of student responses allows for other types of education research.
For example, learning retention can be addressed by examining the performance of 
students who encounter the same problem multiple times.
The left panel of Fig.~\ref{fig:repeatProbs} shows the frequency distribution of multiple 
occurrences of the same problem.
There are nearly 10,000 instances of a student repeating a problem, and a few 
hundred instances of seeing a problem six times.
The center panel shows a marked drop in the median time to solution, by a factor of 
two for the second and third occurrences, constant thereafter at just under 20 seconds.
The correct response fraction grows in turn, climbing from $52\%$ on the first view to 
above 90\% for the fourth and higher occurrences.
 
\subsection{Sustainability and Security} 

The coauthors of this paper are the original developers of PR and its {\em de facto} 
managers.
As PR expands beyond the Physics Department, we must address the management 
and governance of the service explicitly.
We have begun discussions with campus organizations, including instructional 
support staff within the College of Literature, Science and Arts and the U-M Library, 
about migrating server management to a central unit.
Departments would maintain management of the problem libraries and control the 
structure of the topics associated with their courses.
Future development could be governed collectively by a small committee with 
representation from departments and the relevant service units.

Instructional techniques constantly evolve, and the winds of change in higher 
education are blowing harder with the rise of new instruments like massively on-line, 
open courses.
Still, the nearly century-old tradition of timed examination remains a workhorse vehicle to 
assess learning in physics and many other STEM areas.
While new methods of assessment will undoubtedly emerge, it is likely that well-posed, focused questions will continue to play a useful role in learning STEM subjects.
We thus anticipate the need for a service like Problem Roulette into the foreseeable 
future. 

Since the scoring information in PR is not currently incorporated as a component of 
student grade, the security risk posed by storing student performance information is 
relatively low.
The use of a virtual server provided by U-M's central IT organization means that 
system software and network access is maintained by them in a real-time manner.
The server also manages student and staff authentication using CoSign\cite
{cosignSite}.

\section{Summary}\label{sec:summary}

We describe a hybrid cloud service called Problem Roulette that enables random 
study of topical, choose-from-a-list problems.
The server architecture is lightweight, and the main cost of entry is in creating content 
in the form of libraries of individual questions published as web documents.
Our source code for the service is available in the public domain.

Using roughly 1000 problems scraped from previous exams, PR was made available 
as an optional study service to students in introductory physics courses at the 
University of Michigan in Fall 2012.
Usage by nearly 1000 students across four courses spiked in the days leading up to 
scheduled exams, with activity peaking at 800 problems served per hour on the day of 
the exam.

We provide an initial evaluation of the effectiveness of practicing test problems to 
learning physics in the second semester Science/Engineering course.
We find that the population of students who worked more than fifty problems on PR 
during the term (the median value) outperform their complement with the same 
beginning-of-term U-M GPA by an average of 0.22 grade points, well above the noise 
level of 0.04.
The increase is 0.40 grade points when GPA is ignored, reflecting the fact that higher-
GPA students tend to use PR more.

We are in the early stages of expanding the service to other departments and are 
working to develop new management and governance structures with campus 
stakeholders.

\newpage

\appendix*   

\section{Source code and operating environment.}


Source code for PR is hosted at \url{<https://bitbucket.org/mcmills/problemroulette>} 
under a ``BSD-new'' open source license.
The source contains a README.txt file which explains how to configure a server and 
install problem roulette itself.
That file also explains how to fork the code so that beneficial modifications can be 
merged back into the master branch and made available to other forks.  

For specificity, the current version of PR is run using the following software versions:  
MySQL client v.5.1.60;  PHP v.5.2.9; and jQuery v.1.10.1.
The jQuery suite is a JavaScript library that facilitates browser compatibility.

\begin{acknowledgments}

We gratefully acknowledge assistance from Sam Shreeman, Tyler Hughes and 
Mallory Traxler as well as the many U-M Physics faculty members who provided old 
exams.   
We are grateful to the U-M Provost and to Instructional Support Services in the College 
of Literature, Science and Art for funding support. 

\end{acknowledgments}

\end{document}